\documentclass[10pt]{article}
\usepackage{cite}
\usepackage{graphicx,amsmath,amssymb,amsbsy,latexsym,amsthm}
\usepackage{bm}
\usepackage[mathscr]{eucal}

\theoremstyle{plain}

\newcommand{\startproof}{\noindent\textbf{Proof.} }
\newcommand{\finishproof}{\hfill $\blacksquare$ \\}

\def\R{\mathbb{R}}

\newcommand{\eqa}{\begin{eqnarray}}
\newcommand{\neqa}{\end{eqnarray}}
\newcommand{\be}{\begin{equation}}
\newcommand{\ee}{\end{equation}}

\newcommand{\geom}{\mathrm{geom}}
\newcommand{\phys}{\mathrm{phys}}
\newcommand{\Regge}{\mathrm{Regge}}

\newcommand{\sut}[1]{{\bm #1}}

\newcommand{\ors}{\omega}
\newcommand{\plebs}{\nu}

\begin{document}

\title{Corrigendum: The Plebanski sectors of the EPRL vertex\\
\Large 
(2011 \textit{Class. Quant. Grav.} \textbf{28} 225003, \texttt{arXiv:1107.0709}) }

\author{Jonathan Engle\thanks{jonathan.engle@fau.edu}
 \\[1mm]
\normalsize \em Department of Physics, Florida Atlantic University, Boca Raton, Florida, USA}
\date{\today}
\maketitle\vspace{-7mm}

\begin{abstract}
We correct what amounts to a sign error in the proof of part (i.) of theorem 3.  The Plebanski sectors isolated by the linear simplicity constraints do not change --- they are still the three sectors (deg), (II+), and (II-).  What changes is the characterization of the continuum Plebanski two-form corresponding to the first two terms in the asymptotics of the EPRL vertex amplitude for Regge-like boundary data.  These two terms do not correspond to Plebanski sectors (II+) and (II-), but rather to the two possible signs of the \textit{product} of the sign of the sector --- $+1$ for (II+) and $-1$ for (II-) --- and the sign of the orientation
$\epsilon_{IJKL}B^{IJ} \wedge B^{KL}$ determined by $B^{IJ}$.
This is consistent with what one would expect, as this is exactly the sign which classically relates the BF action, in Plebanski sectors (II+) and (II-), to the Einstein-Hilbert action, whose discretization is the Regge action appearing in the asymptotics. 
\end{abstract}

\section*{The error and the corrected final result}

The error lies in part (i.) of theorem 3 of the paper.   In order to state this error, let us define a \textit{numbered 4-simplex}
to be a geometrical 4-simplex with vertices numbered arbitrarily, and each tetrahedron numbered by the vertex it does not contain.  
An `ordered 4- simplex' as defined in definition 3  is then a numbered 4-simplex that additionally satisfies a condition relating 
the numbering to orientation.
In order for the argument for part (i.) of theorem 3 to be valid, the numbered 4-simplex gauranteed by 
the reconstruction theorem must be `ordered', because it is then used to calculate the Plebanski sector of the 
geometrical bivectors, whose well-definition requires this. But, in general, the reconstructed 4-simplex will not be ordered.

This is the error in the paper.  As we will see, it can be easily corrected, and upon correction, the interpretation of the terms in the asymptotics of the vertex amplitude will no longer involve only Plebanski sectors, but also the \textit{orientation}
$\epsilon_{IJKL} B^{IJ} \wedge B^{KL}$ determined by the continuum two-form $B_{\mu\nu}^{IJ}$ reconstructed from the discrete data at the critical points.  Specifically, let 
\begin{displaymath}
\ors(B_{\mu\nu}):={\rm sgn}\left(\mathring{\epsilon}^{\alpha\beta\gamma\delta}
\epsilon_{IJKL} B^{IJ}_{\alpha\beta} B^{KL}_{\gamma\delta}\right)
\end{displaymath}
 where $\mathring{\epsilon}^{\alpha\beta\gamma\delta}$
is the fixed orientation on $M \cong \R^4$, and let $\plebs(B_{\mu\nu}) = +1, -1$ if $B^{IJ}_{\mu\nu}$ is 
in Plebanski sector (II+) or (II-), respectively, and let $\plebs(B_{\mu\nu}) = 0$ otherwise.   Then the first and second terms in the asymptotics of equation (3.10) correspond to 
critical points where $\ors \plebs = +1$ or $-1$, respectively.

Note that this modified result is exactly what one would expect: The first and second terms in equation (3.10) are
respectively $e^{iS_\Regge}$ and $e^{-iS_\Regge}$, where $S_\Regge$ is the Regge action.  The Regge action
is a discretization of the Einstein-Hilbert action $S_{EH}$, and the relation of the BF action $S_{BF}$ to the Einstein-Hilbert action, 
in Plebanski sectors (II+) and (II-), is precisely $S_{BF} = \ors \plebs S_{EH}$.

\section*{Details of the correction}

In the following, $\{B_{ab}\}$ shall always denote a ``\textit{discrete Plebanski field}'' in the sense of \cite{engle2011}
--- that is, a set of $\mathfrak{so}(4)$ algebra elements $B_{ab}^{IJ} = -B_{ab}^{JI}$ satisfying closure
($\sum_{b: b \neq a} B_{ab}^{IJ} = 0$) and orientation ($B_{ab}^{IJ} = - B_{ba}^{IJ}$).  
The algebra indices $IJ$ will usually be suppressed.  
The algebra elements $B_{ab}$ are also referred to as \textit{bivectors}
due to the antisymmetry of the algebra indices.
Let $B_{\mu\nu}(\{B_{ab}\}, \sigma)$ denote the unique
$\mathfrak{so}(4)$-valued two form, constant with respect to $\partial_a$, such that 
its integral over each triangle $\Delta_{ab}(\sigma)$ of the numbered 4-simplex $\sigma$ is equal to the algebra element $B_{ab}$.
The existence and uniqueness of the two-form $B_{\mu\nu}$ satisfying these conditions 
is ensured by 
Lemma 1 of \cite{engle2011}.
The proof of Lemma 1 does not depend on $\sigma$ being ordered; see also the related work in \cite{bfh2009}.
When defining the Plebanski sector and orientation of a set of 
algebra elements $\{B_{ab}\}$, however, we will see that it \textit{is} necessary to restrict $\sigma$ to be ordered, but
for the mere reconstruction of $B_{\mu\nu}$ itself, we can and do omit this restriction.

We begin by noting that the proof of theorem 1 in \cite{engle2011} actually succeeds in proving the following much stronger
statement.

\noindent\textbf{Theorem 1, stronger statement.}
{\it For any numbered 4-simplex $\sigma$, $B_{\mu\nu}(B^\geom_{ab}(\sigma), \sigma)$ is in Plebanski sector (II+) and has orientation 
$\ors = +1$.}

Let us next prove two lemmas, which will make the corrected proof of part (i) of Theorem 3 a single line.
For these two lemmas, let $P$ denote any orientation-reversing diffeomophim such that $P\circ P$ is the identity.

\noindent\textbf{Lemma 3}
{\it Given any discrete Plebanski field $\{B_{ab}\}$ and any numbered 4-simplex $\sigma$, 
\begin{equation}
\label{lemm3eq}
B_{\mu\nu}(\{B_{ab}\}, P\sigma) = - P^* B_{\mu\nu}(\{B_{ab}\},\sigma) . 
\end{equation}
}
{\startproof
As mentioned in \cite{engle2011}, the only background structures used in the construction of the continuum two-form 
$B_{\mu\nu}(\{B_{ab}\}, \sigma)$ are the flat connection $\partial_a$ and the fixed orientation $\mathring{\epsilon}^{\alpha\beta\gamma\delta}$.  We begin by making the fixed orientation an explicit argument in the
construction $B_{\mu\nu}(B_{ab}, \sigma, \mathring{\epsilon})$, so that, 
%
%
given $\{B_{ab}^{IJ}\}$,  
$(\sigma, \mathring{\epsilon}) \mapsto
B_{\mu\nu}^{IJ}$ is covariant under the symmetry group of $\partial_a$, that is, under \textit{all} of $GL(4)$.  In particular,
for $P \in GL(4)$, it follows that
\begin{equation}
\label{Pcov}
B_{\mu\nu}(\{B_{ab}\}, P\sigma, P\mathring{\epsilon}) = P^*B_{\mu\nu}(\{B_{ab}\},\sigma, \mathring{\epsilon}) .
\end{equation}
Furthermore, by definition of the reconstructed two-forms (and introducing the orientation $\mathring{\epsilon}$ as an 
explicit argument also of each oriented triangle $\Delta_{ab}(\sigma, \mathring{\epsilon})$), one has
\begin{eqnarray*}
\nonumber
 \int_{\Delta_{ab}((P\sigma), \mathring{\epsilon})} B(\{B_{a'b'}\}, P \sigma, \mathring{\epsilon})
&:=& B_{ab} =: \int_{\Delta_{ab}((P\sigma), P\mathring{\epsilon})} B(\{B_{a'b'}\}, P \sigma, P\mathring{\epsilon}) \\
&=& - \int_{\Delta_{ab}((P\sigma), \mathring{\epsilon})} B(\{B_{a'b'}\}, P \sigma, P\mathring{\epsilon}) 
=  - \int_{\Delta_{ab}((P\sigma), \mathring{\epsilon})} P^* B(\{B_{a'b'}\}, \sigma, \mathring{\epsilon}) 
\end{eqnarray*}
where the second to last equality holds because the sole effect of replacing $P \mathring{\epsilon}$ with $\mathring{\epsilon}$ in the
argument for triangle $\Delta_{ab}$ is to reverse the orientation of the triangle and hence negate the value of the integral,
and the last equality holds because of (\ref{Pcov}).
Because the continuum two-forms are constant with respect to $\partial_a$ and are completely determined by the
values of the above integrals for all $a,b$ \cite{engle2011}, it follows that the integrands of the first and last expressions are equal, 
which, combined with
$B_{\mu\nu}(\{B_{ab}\}, \sigma) := B_{\mu\nu}(\{B_{ab}\}, \sigma, \mathring{\epsilon})$, implies the claimed result (\ref{lemm3eq}).
\finishproof}

In order to understand the significance of the above lemma, we first note that,
for $B_{\mu\nu}$ in Plebanski sector (II+) or (II-), 
the action of $P$ on $B_{\mu\nu}$ flips the orientation of $B_{\mu\nu}$ while leaving its Plebanksi sector 
unchanged, and negation of $B_{\mu\nu}$ flips its Plebanski sector while leaving its orientation unchanged.
These facts, together with the above lemma imply
\begin{equation}
\label{Pflip}
\ors(B_{\mu\nu}(\{B_{ab}\}), P\sigma) = - \ors(B_{\mu\nu}(\{B_{ab}\}), \sigma)
\quad \text{and} \quad  
\plebs(B_{\mu\nu}(\{B_{ab}\}), P\sigma) = - \plebs(B_{\mu\nu}(\{B_{ab}\}), \sigma).
\end{equation}
Because of the above equations, if we wish to use $B_{\mu\nu}(\{B_{ab}\}, \sigma)$
to define a Plebanski sector and orientation for a given set of algebra elements 
$\{B_{ab}\}$, 
a restriction must be placed on the numbered 4-simplex $\sigma$ such that it not possible
to use both a 4-simplex $\sigma'$ and its parity reversal $P\sigma'$;
otherwise the Plebanski sector and orientation of $\{B_{ab}\}$ will be ill-defined.
The restriction used is precisely that $\sigma$ be \textit{ordered} in the sense of \cite{engle2011}.
Once this restriction is made, $\plebs(B_{\mu\nu}(\{B_{ab}\}), \sigma)$ and 
$\ors(B_{\mu\nu}(\{B_{ab}\}), \sigma)$ are \textit{independent} of the remaining freedom in
$\sigma$. This was proven for $\plebs(B_{\mu\nu}(\{B_{ab}\}), \sigma)$ in Lemma 2 of \cite{engle2011}.
For $\ors(B_{\mu\nu}(\{B_{ab}\}), \sigma)$, the proof follows from the same argument, together with the fact
that,  for any orientation preserving diffeomorphism $\varphi$, $\omega(\varphi^* B_{\mu\nu}) = \omega(B_{\mu\nu})$.
Thus, one may define $\plebs(\{B_{ab}\}):= \plebs(B_{\mu\nu}(\{B_{ab}\}), \sigma)$
and $\ors(\{B_{ab}\}):= \ors(B_{\mu\nu}(\{B_{ab}\}), \sigma)$ where any ordered $\sigma$ may be used.
(The significance of the ordering condition on $\sigma$ in this context is essentially that, by imposing a certain compatibility between the numbering and the orientation, the ordering condition \textit{endows} the numbering of
vertices with orientation information which turns out to be essential in extracting the Plebanski sector
and dynamical orientation from the algebra elements $\{B_{ab}\}$.)

\noindent\textbf{Lemma 5}
{\it
For any numbered 4-simplex $\sigma$,
$\ors\left(\{B_{ab}^\geom(\sigma)\}\right)\plebs\left(\{B_{ab}^\geom(\sigma)\}\right) = 1$.
}

{\startproof  

\textit{Case 1, $\sigma$ is ordered:} 
Then $\plebs(\{B_{ab}^\geom(\sigma)\}):= \plebs(B_{\mu\nu}(\{B_{ab}^\geom(\sigma)\}, \sigma))
= + 1$ and $\ors(\{B_{ab}^\geom(\sigma)\}):= \ors(B_{\mu\nu}(\{B_{ab}^\geom(\sigma)\}, \sigma))
= + 1$ where, in each of these equations, the first equality follows by definition and the final equality is implied
by the above stronger version of theorem 1.

\textit{Case 2, $\sigma$ is not ordered:}  Then $P\sigma$ is an ordered 4-simplex, 
so that 
\begin{displaymath}
\plebs\left(\{B_{ab}^\geom (\sigma)\}\right) :=
\plebs\left(B_{\mu\nu}(\{B_{ab}^\geom(\sigma)\}, P\sigma) \right)
= - \plebs\left(B_{\mu\nu}(\{B_{ab}^\geom(\sigma)\}, \sigma) \right)
= - 1
\end{displaymath}
and
\begin{displaymath}
\ors\left(\{B_{ab}^\geom (\sigma)\}\right) :=
\ors\left(B_{\mu\nu}(\{B_{ab}^\geom(\sigma)\}, P\sigma) \right)
= - \ors\left(B_{\mu\nu}(\{B_{ab}^\geom(\sigma)\}, \sigma) \right)
= - 1
\end{displaymath}
where, in each of the above equations, the first equality follows by definition, the second equality follows
from equation (\ref{Pflip}), and the final equality is implied by the above stronger version of theorem 1.
%
%

In both cases, one has $\ors\left(B_{ab}^\geom (\sigma)\right) \plebs\left(B_{ab}^\geom (\sigma)\right) = 1$,
as claimed.
\finishproof}

The corrected statement and proof of part (i.) of theorem 3 are then as follows.

\noindent\textbf{Theorem 3, part (i), corrected.}
{\it
Suppose $\{A_{ab}, \sut{n}_{ab}\}$ is a set of non-degenerate reduced boundary data
satisfying closure and $\{X_a^\pm\}$ are such that orientation is satisfied.
If $\{X^-_a\} \not\sim \{X^+_a\}$, then $\{B^{\phys}_{ab}(A_{ab}, \sut{n}_{ab}, X^\pm_a)\}$ 
is either in Plebanski sector (II+) or (II-). Furthermore, the 
sign $\mu$ in the reconstruction theorem equals $\ors \plebs$.
}

{\startproof
Let $\sigma$ denote the numbered 4-simplex gauranteed by the reconstruction theorem, unique upto translation, rotation, and inversion. Using the relation $B_{ab}^\phys = \mu B_{ab}^\geom(\sigma)$ 
between the physical and geometrical bivectors in the reconstruction theorem, and using
Lemma 5, one has
\begin{displaymath}
\ors\left(B_{ab}^\phys\right) \plebs\left(B_{ab}^\phys\right)
= \ors\left(B_{ab}^\geom(\sigma)\right) \left( \mu \cdot \plebs\left(B_{ab}^\geom(\sigma)\right) \right)
= \mu .
\end{displaymath}
\finishproof}

\section*{Acknowledgements}
The author would like to thank the Institute for Quantum Gravity at the University of Erlangen-Nuernberg 
for an invitation to give a seminar on this topic, 
the preparation for which led to the realization of the error corrected here, and would like to thank
Carlo Rovelli and John Barrett for emphasizing the importance of this result and the importance of publishing its correction
quickly. This work was supported in part by 
NSF grant  PHY-1237510
and by NASA 
through the University of Central Florida's NASA-Florida Space Grant Consortium.

%

\begin{thebibliography}{1}

\bibitem{engle2011}
J.~Engle, ``{The Plebanski sectors of the EPRL vertex},'' {\em Class. Quant.
  Grav.}, vol.~28, p.~225003, 2011.


\bibitem{bfh2009}
J.~Barrett, W.~Fairbairn, and F.~Hellmann, ``Quantum gravity asymptotics from
  the {SU}(2) 15j symbol,'' {\em Int. J. Mod. Phys. A}, vol.~25,
  pp.~2897--2916, 2010.

\end{thebibliography}
%
%
%

\end{document}